\documentstyle[epsfig, subfigure]{mn}

\def\ltapprox{\raise 2pt \hbox {$<$} \kern-1.1em \lower 5pt \hbox {$\approx$}}
\def\ltsim{\raise 2pt \hbox {$<$} \kern-1.1em \lower 4pt \hbox {$\sim$}}
\def\gtsim{\raise 2pt \hbox {$>$} \kern-1.1em \lower 4pt \hbox {$\sim$}}

\title{Particle reacceleration by compressible turbulence in galaxy
clusters: effects of reduced mean free path}

\author[G. Brunetti, A. Lazarian]
      {G. Brunetti,$^1$ 
       A. Lazarian$^2$\\
       $^1$ INAF/Istituto di Radioastronomia, via Gobetti 101,
       I--40129 Bologna, Italy \\
       $^2$ Department of Astronomy, University of Wisconsin at Madison,
       5534 Sterling Hall, 475 North Charter Street, Madison,
       WI 53706, USA\\
}

\begin{document}
\maketitle

\begin{abstract}
Direct evidence for in situ particle acceleration mechanisms
in the inter-galactic-medium (IGM) is provided by the diffuse Mpc--scale 
synchrotron emissions observed from galaxy clusters.
It has been proposed that MHD turbulence, generated during cluster-cluster  
mergers, may be a source of particle reacceleration in the IGM.
Calculations 
of turbulent acceleration must account self-consistently for
the complex non--linear coupling between turbulent waves and particles.
This has been calculated in some detail under the assumption that turbulence
interacts in a collisionless way with the IGM.
In this paper we explore a different picture of 
acceleration by compressible
turbulence in galaxy clusters, where the interaction between turbulence
and the IGM is mediated by plasma instabilities and maintained collisional
at scales much smaller than the Coulomb mean free path.
In this regime most of the energy of fast modes is channelled into
the reacceleration of relativistic particles and the acceleration process
approaches a universal behaviour being self-regulated by the back-reaction 
of the accelerated particles on turbulence itself.
Assuming that relativistic protons contribute to several percent (or 
less) of the cluster energy, consistent with the FERMI observations of nearby 
clusters, we find that compressible turbulence at the level of 
a few percent of the thermal energy can reaccelerate relativistic
electrons at GeV energies, that are necessary to explain the observed diffuse
radio emission in the form of giant radio halos.
\end{abstract}


\maketitle

\section{Introduction}

Mergers between galaxy clusters are the most energetic events in the present 
Universe.
During these collisions a fraction of the gravitational binding--energy 
of massive Dark Matter halos can be channelled into shocks and turbulence
that may accelerate relativistic protons and electrons (e.g. Ryu et al
2003; Cassano \& Brunetti 2005; Brunetti \& Lazarian 2007;
Hoeft \& Br\"uggen 2007; Pfrommer et al 2008; 
Skillman et al 2008; Vazza et al 2009),
while collisions between
the accelerated protons and the thermal protons generate secondary particles
(e.g. Blasi \& Colafrancesco 1999; Pfrommer \& Ensslin 2004).
This makes galaxy clusters unique laboratories to study particle 
acceleration in diluted astrophysical plasma.

\noindent
Radio observations of galaxy clusters probe these complex processes through 
the study of cluster--scale synchrotron emission generated 
by relativistic electrons that gyrate in the magnetic fields
of the IGM.
Giant radio halos are the most spectacular, and best studied, 
examples of cluster--scale synchrotron sources. 
They are steep--spectrum, low brightness diffuse emissions that 
extend similarly to the hot X--ray emitting gas 
(eg. Ferrari et al 2008 for a review) and that are found in
merging clusters (eg. Cassano et al 2010 and ref therein). 
The morphological and spectral properties of a number of radio
halos suggest that the emitting electrons are accelerated by spatially 
distributed and "gentle" (i.e. poorly efficient, with acceleration
time $\sim 10^8$yrs) mechanisms 
(e.g. Brunetti et al 2008).

A model put forward for the origin of giant radio halos assumes that 
relativistic particles in the IGM are reaccelerated by MHD turbulence that is
generated during massive cluster--cluster mergers 
(e.g. Brunetti et al. 2001, 2004; 
Petrosian 2001; Fujita et al 2003; Cassano \& Brunetti 2005). 
The theory of MHD turbulence seriously advanced in the last decades 
(see Cho, Lazarain \& Vishniac 2003 for a review), affecting 
our view of particle acceleration in astrophysical plasmas (e.g. Chandran 2000;
Yan \& Lazarian 2002; 2004, 2008).
Recently (Brunetti \& Lazarian 2007, 2010) we considered these advances
to develop a comprehensive picture of turbulence in the IGM and 
to study stochastic reacceleration of relativistic particles.
We suggested that compressible MHD turbulence (essentially fast modes), 
generated during energetic cluster--mergers, provides the most important
driver of stochastic particle reacceleration in the IGM, and that the 
interaction between 
this turbulence and the relativistic electrons may explain the origin of 
radio halos,  
provided that a fraction (10-25 \%) of the energy 
dissipated 
during mergers is channelled into these modes.

\noindent
To what extend these calculations were accurate depends on our understanding 
of the properties of IGM turbulence.
The interaction between fast modes and both the 
thermal IGM and the relativistic particles was 
assumed collisionless, in which case about 90 percent of the
energy of fast modes goes into heating of the thermal plasma, while only 
10 percent is available for the reacceleration of relativistic 
particles (eg. Brunetti \& Lazarian 2007); in the following we refer
to this assumption as {\it collisionless} IGM.

\noindent
On the other hand one may think about a different picture.
Indeed it can be argued that the degree of collisionality of astrophysical 
plasmas is underestimated when only Coloumb collisions are taken 
into account (eg. Lazarian et al 2010).
Instabilities are naturally generated into the IGM 
(eg. Schekochihin et al 2005) and make the fluid more collisional, 
in which case a larger fraction of the energy of compressible MHD turbulence 
may become available for the reacceleration of relativistic particles.

\noindent
At the same time 
the maximum energy budget available for cosmic rays
in the IGM can be efficiently constrained from
the recent upper limits to the gamma ray emission from 
nearby galaxy clusters (eg. Aharonian et al.~2009;
Ackermann et al 2010) and to the Mpc-scale 
radio emission in clusters without radio halos (eg. Brunetti et al 2007),
and this provide important information for theoretical models.

Consequently in this paper we explore the process of particle reacceleration
by compressible turbulence in the IGM by assuming a picture
of turbulence different from that considered in previous studies.
According to this picture 
plasma instabilities are naturally generated in the IGM and decrease the 
effective collisional scale of the thermal IGM making the interaction
between turbulence and the thermal plasma more collisional.

\noindent
In Sect.~2 we discuss the properties of MHD turbulence and the
effect of reduced mean free path in determining the way the 
IGM interacts with turbulence.
In Sect.3 we discuss consequences of a reduced mean free path
in the IGM on turbulence damping and
on the reacceleration of relativistic particles;  
in Sect.4 we report on the case of the 
reacceleration of relativistic protons and of secondary electrons.
In Sect.5 and 6 we provide a more general discussion and our 
conclusions, respectively.

\section{Turbulence in galaxy clusters}
\label{sec:sample}

\subsection{Effective collisionality of the IGM}

It is well known that the mean free path of thermal protons due to 
Coulomb collisions in the hot IGM is very large, 
ten to hundred kpc (e.g. Sarazin 1986).
Fluids in such a collisionless regime can be very different from their 
collisional counterparts (Schekochihin et al. 2005; 2010).
Several instabilities (e.g. firehose, mirror,
gyroresonance etc.) can be generated in 
the IGM in the presence of turbulence, leading to a 
transfer of the energy of
large-scale compressions to perturbations on smaller scales. 

Many instabilities have growth rate which peaks at scales
near the particle gyroradius,
making very large the scale separation between the energy injection scale 
and the scale where this energy is being deposited.
On one hand scatterings induced by instabilities dramatically
{\it reduce the effective mean free path} of thermal ions (e.g. 
Schekochihin \&
Cowley 2006) {\it decreasing the effective viscosity} of the IGM, 
at the same time these scatterings 
may change the {\it effective collisionality} of the plasma.
Indeed, the usual notion of collisions in plasmas assumes 
Coulomb collisions. However, charged particles can be randomized if 
they interact with perturbed magnetic field. If this field is a result 
of plasma instabilities, {\it the process can be viewed as
the collective interaction of an individual ion with the rest of the 
plasma}, which is the process mediated by magnetic field. 
As a result, the fluid would behave as collisional on scales less that 
the Coulomb mean free path. 
This issue has been addressed in Lazarian \& Beresnyak (2006) 
for the case of a collisionless fluid subject to the 
gyroresonance instability that is driven by the
anisotropy of the particle distribution in the momentum space
that arises from magnetic field compression; the larger the magnetic 
field compression, the higher the anisotropy induced and the higher is 
the instability growth rate. 
They found that the turbulent magnetic compressions on the scale of the
mean free path and less are the most effective for inducing the 
instability\footnote{The larger scale compressions do still induce the 
instability, but 
their effect is reduced due to their reduced ability to induce large 
changes of $B$ over the time scale between scattering.}.
As the scattering happens on magnetic 
perturbations induced by the instability, the mean free path of 
particles decreases as a result of the operation of the instability. 
This results in the process being self-regulating, i.e. the stronger 
the turbulence at the scale of injection, the smaller is the mean free 
path of plasma particles and the larger is the span of scales over 
which the fluid behaves as essentially collisional. 

\noindent
Given these general considerations, 
in the following we shall assume that 
the interaction between the turbulent modes and the thermal IGM 
is similar to that of {\it collisional fluids}
on scales which are {\it less than the Coulomb mean free path}
but {\it larger than the mean free path arising from particle scattering 
by magnetic perturbations driven by instabilities}; in the following we
refer to this assumption as {\it collisional} IGM.

\subsection{The turbulent picture in the IGM}

At large scales turbulence in the IGM is 
likely super-Alfvenic, the injection velocity 
$V_L$ being substantially greater that the 
Alfven velocity $v_A$, in this case turbulence in the IGM behaves as 
hydro--turbulence (see Lazarian 2006, Brunetti \& Lazarian 2007).
In the Kolmogorov cascade the turbulent velocity $V_l$ scales as 
$V_L (l/L_o)^{1/3}$, and at scales less than the transitional scale,
$l_A\sim L_o (V_L/v_A)^{-3}$, 
turbulence gets sub-Alfvenic and obeys the MHD turbulence relations
(see Goldreich \& Srindhar 1995, and also Lazarian \& Vishniac 1999, 
Cho \& Lazarian 2003)\footnote{MHD turbulence theory has a long history 
(see Biskamp 2003) and its details are
still a subject of hot debates. 
However, recent numerical calculations are  roughly consistent with the model 
of strong Alfvenic turbulence in Goldreich \& Sridhar (1995) (see 
Beresnyak \& Lazarian 2009) and confirm scaling of compressible modes reported 
in Cho \& Lazarian (2003) (see Kowal \& Lazarian 2010). } 
down to collisionless scales. Note, that for the IGM the scale of the 
transition is $l_A \sim 0.1-1$ kpc (e.g. Brunetti \& Lazarian 2007).

\noindent 
In the range of scales where the interaction between turbulence
and the thermal IGM is {\it collisional} 
the most important damping of turbulent motions is due to relativistic 
particles.
This is particularly important for fast 
modes (Brunetti \& Lazarian 2007), while it is well known that the damping 
of solenoidal 
and slow modes components of the turbulence is much less efficient, 
at least at relatively large scales (e.g. Yan \& Lazarian 2004).
In the MHD-- regime, $l \leq l_A$, MHD numerical simulations have
shown that a solenoidal turbulent forcing gets
the ratio between the amplitude of Alfv\'en $\approx \delta V_s$ and 
fast $\approx \delta V_c$ modes in the form (Cho \& Lazarian 2003) :

\begin{equation}
{{(\delta V)_c^2}\over{(\delta V)_s^2}} \sim
{{(\delta V)_s v_A }\over{ c_s^2 + v_A^2 }} 
\label{transfercl03} 
\end{equation} 

\noindent
which essentially means that coupling between these two
modes is inefficient at the sufficiently small scales where the perturbations 
are sub-Alfvenic. This allows
us to talk about separate cascades of fast, slow and Alfven modes in 
agreement with the simulations which performed
mode separation and studied those cascades (Cho \& Lazarian 2003, 
Kowal \& Lazarian 2010). 

\noindent
Consequently we can assume that in the clusters of galaxies the energy 
transported by the cascade 
of fast modes is mainly channelled into the reacceleration of 
relativistic particles, while the compressions of magnetic field arising from 
slow modes 
transports the energy from large to small collisionless scales, 
sustaining the generation of the same 
compressible instabilities that increase the effective collisionality of the 
thermal IGM at
scales smaller than the Coulomb scale.
In this respect, to provide a more quantitative view 
of our picture as an example we adopt the reference 
case of firehose instability (e.g. Chandrasekhar et al.~1958;
Barnes 1966).
The threshold condition for the instability to occur 
with thermal electrons is $\delta T_{\Vert} / T > 1/\beta_{pl}$ 
so that this instability is expected to naturally develop in high beta
plasmas, like the IGM.
The cascading slow modes compress the plasma along the field lines generating 
anisotropies (due to conservation of adiabatic invariant)
in the phase--distribution of thermal particles and 
potentially may drive firehose instability in the IGM. 
Their magnetic--field 
compression factor is (Cho \& Lazarian 2003):

\begin{equation}
{{\delta B }\over{ B }} \sim {{\delta T_{\Vert} }\over{ T }}
\approx {{(\delta V_l)_s }\over{ v_A }} \approx 
\left( {{l}\over{ L_o }} \right)^{1/3} (V_L/v_A)_s
\label{firehose}
\end{equation}

\noindent
(where $(V_L/c_s)_s$ is the Mach number of slow modes)
that ``potentially'' implies a collisional scale 
of thermal IGM $l_f \sim 3 \times 10^{-8} L_o$\footnote{This assumes
that the collisional scale is maintained by the instability
at the minimum scale where instability occurs.} when combined 
with the aforementioned threshold condition for the instability to occur;
this is $l_f \approx 10^{-7}$ times the classical Coulomb mean free path of
the thermal IGM (assuming $L_o \approx 200-300$ kpc).

\section{Dampings of fast modes and particle acceleration}
\label{sec:morph}

Having motivated our picture of a {\it collisional} 
IGM, in this Section we discuss consequences on  
reacceleration of relativistic particles by fast modes.

We consider the most simple situation where turbulence is injected 
at a single scale, with wavenumber $k_o$, and assume that
a fraction of the 
turbulent energy--flux is channelled into fast modes, with rate 
$I^f(k,t) = I^f_o \delta(k-k_o)$.
Under these conditions 
fast modes can be assumed isotropic (e.g. Cho \& Lazarian 2003) with 
quasi--stationary spectrum (Brunetti \& Lazarian 2007, 2010) :

\begin{equation}
{\cal W}^f(k) \approx c_w
\left(
I^f_o \rho \langle V_{\rm ph}\rangle \right)^{ {1\over 2} }
k^{- {3 \over 2} }
\label{cascade_spectrum}
\end{equation}

\noindent
for $k_o < k < k_{c}$, where $c_w$ is of the order of unity
and $k_c$ is the cut--off scale where collisionless  
dampings become more efficient than the process of
wave--wave cascading.

\noindent
Under the hypothesis discussed in the previous Section, 
the most important
collisionless damping of fast modes is due to 
the Transit-Time-Damping (TTD) resonance with relativistic particles  
(eg. Schlickeiser \& Miller 1998; Yan \& Lazarian 2004;
Brunetti \& Lazarian 2007, 2010):

\begin{equation}
\Gamma_{CR}
\approx - {{\pi^2}\over{8}} k
\langle {{ |B_k|^2}\over{ {{\cal W}}^f }} \rangle
{{ \sin^2\theta}\over{| \cos\theta|}}
{{ c_{\rm s}^2 }\over{B_o^2}}
\int p^4 dp {{ \partial f(p)}\over{\partial p}} 
\label{GammaCR}
\end{equation}

\noindent
where $B_o$ is the background (unperturbed) magnetic field, $c_s$ is the
sound speed and 
$|B_k|^2/{{\cal W}^f}$ is the ratio between magnetic field 
fluctuations and total energy in the mode (the quantity $\langle .. \rangle$
indicates average with respect to the angle between mode
wavenumber and the background magnetic field).
The cut--off scale is\footnote{provided that $k_c \leq k_{coll}$,
$k_{coll}$ being the wavenumber where the IGM becomes collisionless} :

\begin{equation}
k_{c} = k_{c}^{CR}= c_k
{{I^f_o}\over{\rho  c_{\rm s} }}
\left( {{\langle \Gamma_{CR}(k,\theta) \rangle}\over{k}} \right)^{-2}
\label{kd}
\end{equation}

\noindent
where ${c_k} \sim$ a few 
(Brunetti \& Lazarian 2007; see also Matthaeus \& Zhou 1989,
for details on Kraichnan constants).

\noindent
In this case all the energy of fast modes is channelled into the 
reacceleration of relativistic particles.
The particle--diffusion coefficient in the momentum space due to TTD
is obtained combining e.g. Eq.~47 in Brunetti \& Lazarian (2007) and 
Eq.~\ref{kd} :

\begin{equation}
D_{pp} \simeq
2 c_w c_k^{1/2} 
{{ p^2 I_o^f }\over
{\sum_{e,p} \int dp p^2 c \left|
{{\partial N}\over{\partial p}} -2 {{N}\over{p}}
\right| }} 
\label{dpp1}
\end{equation}

\noindent
where, assuming 
$I_o^f \phi \tau >> \epsilon_{CR}^o$ ($\epsilon_{CR}^o$ the initial
energy density of cosmic rays), it is\footnote{we assume
that damping is dominated by a single species of
relativistic particles}:

\begin{equation}
\int dp p^2 c \left|
{{\partial N}\over{\partial p}} -2 {{N}\over{p}}
\right| \sim c_N \epsilon_{CR} \sim I_o^f \phi c_N \tau
\label{denom}
\end{equation}

\noindent
$c_N$ is a numerical factor that depends on the shape of the 
spectrum of the accelerated particles\footnote{for instance
$c_N = (s+2)$ assuming $N(p) \propto p^{-s}$}.
$\phi$ accounts for the ``intermittent'' nature of the turbulence--injection
process (i.e., $I_o^f \phi \tau$ is the energy density
injected into fast modes in 
the period of time $\tau$), we indeed expect that many
patches of large--scale turbulence can be injected at different times
in a Mpc$^3$ region during a merger.

\noindent
From Eqs.\ref{dpp1} and \ref{denom}  
the process of particle reacceleration is self--regulated by the 
damping of the modes due to the reaccelerated particles.

\noindent
This is a new regime of particle reacceleration by compressible
turbulence. Indeed it greatly differs from 
the case of a {\it collisionless} IGM, where the damping of the modes
is dominated by the collisionless damping with the thermal electrons
in the IGM (e.g. Cassano \& Brunetti 2005; Brunetti \& Lazarian 2007,
2010).
On the other hand, a similar effect, 
the {\it proton--wave boiler}, was observed in 
the case of reacceleration by a hypothetical spectrum of isotropic 
Alfvenic waves (Brunetti et al 2004) where indeed the damping
of the modes was dominated by gyro--resonance with relativistic
protons.

\begin{figure}
\includegraphics[width=0.417\textwidth]{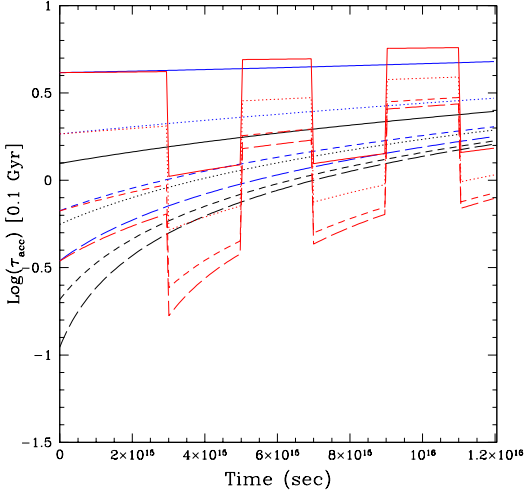}
\vskip -1.5truecm
\caption[]{
The systematic acceleration time--scale is reported as
a function of time.
The initial energy density of cosmic rays is $\epsilon^o_{CR}=0.3$
(black lines) and $3\%$ (blue lines) of the thermal IGM.
Calculations are reported assuming $(V_L/c_s)^2=$0.02 (long--dashed
lines), 0.03 (dashed lines), 0.05 (dotted lines) and 0.07 (solid
lines).
Red lines show a case of intermittent turbulence where we
assume $\epsilon^o_{CR}=0.03 \epsilon_{th}$ and, for simplicity,
we assume that the
turbulent energy varies with time $0.25 \times (V_L/c_s)^2$ and 
$(V_L/c_s)^2$ (same line--code given above).
All calculations assume a spectral index of cosmic rays $\delta=2.6$,
a IGM number density $n_{th}=7 \cdot 10^{-4}$cm$^{-3}$
and temperature $T=10^8$ K.}
\label{fig:substructures}
\end{figure}

\begin{figure*}
\includegraphics[width=0.44\textwidth]{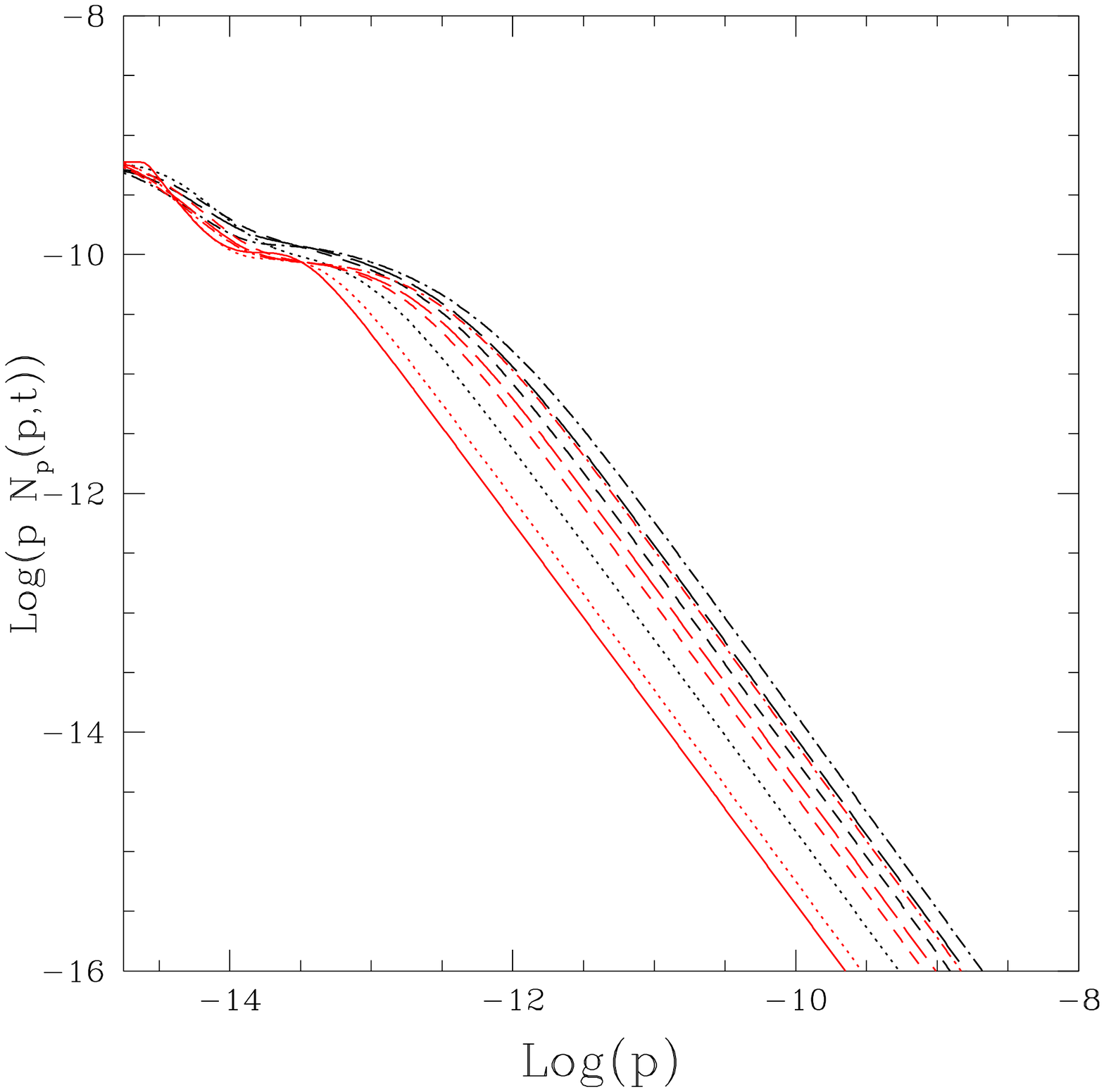}
\includegraphics[width=0.44\textwidth]{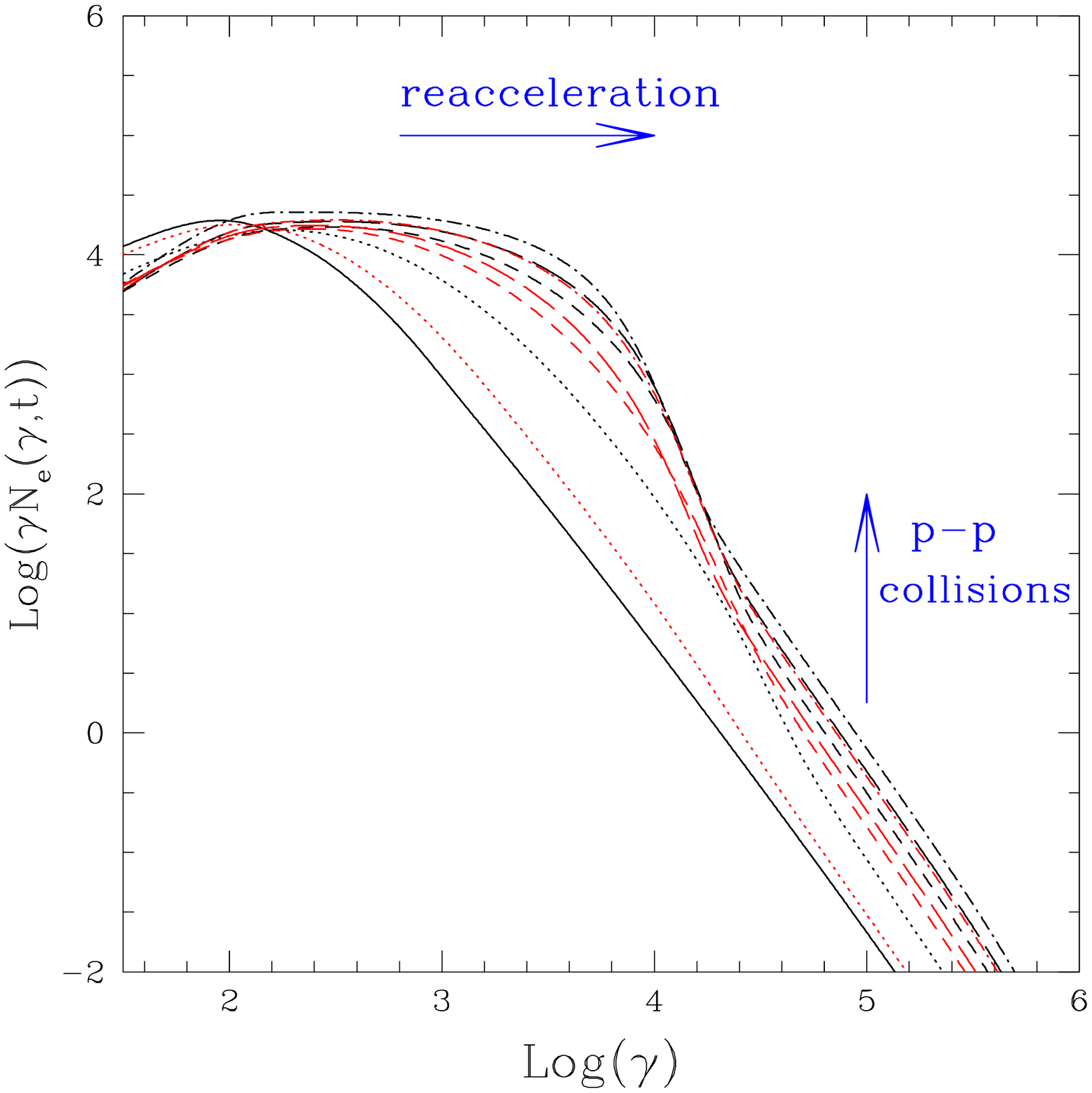}
\vskip -1.5truecm
\caption[]{The evolution with time of the spectrum of relativistic 
protons ({\bf Left Panel}) and of (secondary) electrons ({\bf Right Panel}).
Results are reported for time $\tau =$ 0 (solid lines), 
3 (dotted lines), 9 (short--dashed lines), 12 (long--dashed lines), 
16 $\times 10^{15}$sec (dot--dashed lines). 
Calculations are obtained assuming $\epsilon^o_{CR}=1\%$ of the thermal 
energy, and $(V_L/c_s)^2= 0.07$ (black lines) or intermittent turbulence 
(red lines) (according to Figure 1, with $1$ 
and $0.25\times$ $(V_L/c_s)^2=0.07$). 
Thermal parameters are the same of Figure 1, a magnetic field 
$B_o = 2.5 \mu$G and redshift =0.15 are assumed.} 
\label{fig:substructures}
\end{figure*}

\noindent
According to 
Eqs.~\ref{dpp1} and \ref{denom} we can identify 
two asymptotic regimes of reacceleration ($\phi = 1$ for simplicity) : 
(i) a {\it fast regime}
for $I_o^f \tau_f \sim \rho V_L^2 >> \epsilon_{CR}^o$, 
where $\tau_f$ is the cascading
time of fast modes injected at large scales, and
(ii) a {\it slow regime} for $I_o^f \tau_f << \epsilon_{CR}^o$.

\noindent
In the {\it fast} regime the damping due to relativistic
particles is initially small 
making the acceleration efficiency large.
Under these conditions relativistic
protons rapidly gain energy with the consequence that 
the damping of the turbulent modes by these protons 
increases with time and makes the reacceleration process less
efficient.
After few acceleration times, as soon as 
$\epsilon_{CR} \sim I_o^f \tau$, we expect that the 
process approaches a ``asymptotic'', universal, regime that
does not depend on initial conditions (provided that the
injection rate of turbulence is constant).
In this case the particle--diffusion coefficient in 
the momentum space reads (from Eqs.~\ref{dpp1} and \ref{denom}):

\begin{equation}
D_{pp} \sim 2 {{c_w c_k^{1/2}}\over{c_N \phi}}
\tau^{-1}
\label{dpp2}
\end{equation}

\noindent
i.e. the reacceleration efficiency approaches a universal 
behaviour and decreases (linearly) with time.

\noindent
On the other hand, in the {\it slow regime} relativistic
protons do not increase significantly their energy 
implying a quasi--constant damping of the 
modes; in this case $D_{pp} \propto I_o^f$.

\noindent
All these effects can be seen in Figure 1, where we report the
evolution with time of the systematic reacceleration time,
$\tau_{acc} = p^{2}/(4 D_{pp})$, assuming different initial values
of $I_o^f \tau_f /\epsilon_{CR}^o$ (for simplicity the thermal
energy density and $I_o^f$ are assumed constant with time).
For small $\rho V_L^2/\epsilon_{CR}^o$ we are in the {\it slow
regime} where the acceleration time
does not change significantly with time.
On the other hand, for large $\rho V_L^2/\epsilon_{CR}^o$, we are in the
{\it fast regime} where
initially the acceleration time rapidly increases 
and then approaches the ``asymptotic'' behaviour at later stages.
By adopting a more realistic picture where both the thermal energy
density of the IGM and $I_o^f$ may increase with time during
a merger, or
where turbulence is intermittent (Fig.~1) we expect a less
rapid approach to this ``asymptotic'' behaviour.

\noindent
We believe that the most important consequence of this new regime 
of particle reacceleration by compressible turbulence is the universal
acceleration time--scale, $\sim 10^8$ yrs, that occurs 
when {\it turbulence and cosmic rays reach approximate equipartition}.

\section{Reacceleration of protons and
secondary electrons}

As an example we calculate the evolution with time of the
spectrum of relativistic electrons and protons subject to reacceleration
by fast modes assuming a {\it collisional} IGM.
As a simplification 
we consider only primary 
protons and the secondary electrons produced by inelastic collisions 
between these protons and the IGM.
In this case the damping of the modes is largely dominated by that
with relativistic protons.
This also allows for a prompt comparison with similar calculations
developed under the assumption of {\it collisionless} IGM (Brunetti \&
Lazarian 2010).

\noindent
We model the time evolution of the spectral energy 
distribution of electrons, $N_e$,
with an isotropic Fokker-Planck
equation :

\begin{eqnarray}
{{\partial N_e(p,t)}\over{\partial t}}=
{{\partial }\over{\partial p}}
\Big[
N_e(p,t)\Big(
\left|{{dp}\over{dt}}_{\rm r}\right| -
{1\over{p^2}}{{\partial }\over{\partial p}}(p^2 D_{\rm pp})
\nonumber\\
+ \left|{{dp}\over{dt}}_{\rm i}
\right| \Big)\Big]
+ {{\partial^2 }\over{\partial p^2}}
\left[
D_{\rm pp} N_e(p,t) \right] \nonumber\\
+ Q_e[p,t;N_p(p,t)] \, ,
\label{elettroni}
\end{eqnarray}

\noindent
where $|dp/dt|$ marks radiative (r) and Coulomb (i) losses experienced
by relativistic electrons in the IGM (see Brunetti
\& Lazarian 2010 and ref therein for the relevant formulae),
$D_{pp}$ is the electron diffusion coefficient in the momentum
space due to the coupling with fast modes (Eq.\ref{dpp1}), 
and the term $Q_{e}$ accounts for the injection rate of secondary 
electrons due to p-p collisions in the IGM (following Brunetti \&
Lazarian 2010).

\noindent
Similarly the time evolution of the spectral energy
distribution of protons, $N_p$, is given by :

\begin{eqnarray}
{{\partial N_p(p,t)}\over{\partial t}}=
{{\partial }\over{\partial p}}
\Big[
N_p(p,t)\Big( \left|{{dp}\over{dt}}_{\rm i}\right|
-{1\over{p^2}}{{\partial }\over{\partial p}}(p^2 D_{\rm pp})
\Big)\Big]
\nonumber\\
+ {{\partial^2 }\over{\partial p^2}}
\left[ D_{\rm pp} N_p(p,t) \right] - {{N_p(p,t)}\over{\tau_{pp}(p)}}
\, ,
\label{protoni}
\end{eqnarray}

\noindent
where $|dp/dt_i|$ marks Coulomb losses and $\tau_{pp}$ is the proton 
life--time due to pp collisions in the IGM (see Brunetti \& Lazarian 2010 
and ref therein), and 
$D_{pp}$ is the diffusion coefficient in the momentum
space of protons due to the coupling with fast 
modes (Eq.\ref{dpp1}).

\noindent
The spectral shape of relativistic particles and its evolution with
time is expected to be similar to that in the case of turbulent 
reacceleration in models that assume a {\it collisionless} IGM.
Relativistic protons do not experience relevant energy losses 
in the IGM and TTD resonance in the IGM may reaccelerate supra-thermal 
protons up to high energies.
On the other hand, the case of electrons is more complex.
Radiative losses (due to synchrotron and inverse Compton emission) experienced
by relativistic electrons in the IGM are expected to
prevent an efficient reacceleration of these particles 
above a maximum energy, $\gamma_{max}$. Consequently,
at higher energies the evolution of the electron spectrum is 
basically driven
by the process of injection of fresh electrons due to p--p collisions
(and on its interplay with radiative losses).

\noindent
Results are reported in Figure 2.
Calculations are carried out considering that 
cosmic ray protons contribute to a few percent of the thermal cluster
energy, consistent with the recent limits derived from FERMI observations
of nearby clusters (Aharonian et al.~2009; Ackermann et al 2010).
Under the hypothesis of {\it collisional} IGM, the bulk of turbulence 
(fast modes) is channelled into
reacceleration of cosmic rays and the FERMI limits put corresponding 
constraints 
also on the energy density of turbulent motions in the IGM, about
$(V_L/c_s)^2 \leq 10\%$.
Fig. 2 shows that 
relativistic (secondary) electrons can be reaccelerated at energies of 
several GeV even by assuming that fast modes contribute to only a 
few percent of the energy density of the IGM.
On the other hand
a larger amount of compressible turbulence
is typically requested assuming a {\it collisionless} IGM 
(e.g. Brunetti \& Lazarian 2010 and ref therein).
This however does not imply that relativistic electrons can be
reaccelerated for long periods 
at energies much larger than those in 
the {\it collisionless} case. Indeed larger injection rates of turbulence 
do not make the reacceleration process substantially more efficient,
due to the damping by the relativistic protons that self-regulates
the acceleration efficiency in a few acceleration times.
The effect of damping is also visible in Figure 2 : at later
stages the acceleration of cosmic rays starts to saturate.

\noindent
It is important ti stress that 
self-regulation depends on $I_o^f \phi \tau/\epsilon_{CR}$, thus 
the effect of proton back--reaction on the acceleration efficiency 
becomes less important in the (more realistic)
case of intermittent (or patchy) 
turbulence ($\phi < 1$, Fig.~2) that implies that less energy
in channelled into cosmic ray protons; in this case larger acceleration 
efficiencies are maintained for longer periods of
time (see also Fig.~1).

\section{Discussion}

Whether the thermal 
IGM is {\it collisional} or {\it collisionless} at scales smaller than the
Coulomb scale depends on the effect of reduced mean free path that
is mediated by the plasma instabilities.
Consequently 
the way compressible
turbulence is damped and particle are reaccelerated in the
IGM depends on the interplay between several reference
scales (wavenumbers):  
the collisionless scale, $k_{coll}$, the Coulomb scale, $k_C$,
the turbulence cut--off scale due to collisionless damping with thermal
particles, $k_c^{th}$, and that due to collisionless damping with
relativistic particles, $k_c^{CR}$ (Figure 3 for a sketch of both 
{\it collisional} and {\it collisionless} cases).

The cut--off scale due to collisionless damping with thermal
particles is (Brunetti \& Lazarian 2007) :

\begin{equation}
k_c^{th} \sim {{\cal C}}^{th} k_o \left( {{V_L}\over{c_s}} \right)^4
\label{kcth}
\end{equation}

\noindent
(${\cal C}^{th}$ is a constant) 
while that due to collisionless damping with relativistic
particles (from Eqs.\ref{GammaCR} and \ref{kd}) is :

\begin{equation}
k_c^{CR} \sim {{\cal C}}^{CR} k_o \left( {{V_L}\over{c_s}} \right)^4
\left( {{\epsilon_{CR}}\over{\epsilon_{th}}} c_s \right)^{-2}
\label{kccr}
\end{equation}

\noindent
(${\cal C}^{CR}$ is a constant) 
where $k_c^{CR} >> k_c^{th}$ in the IGM (Brunetti \& Lazarian 2007).

\begin{figure*}
\includegraphics[width=0.407\textwidth]{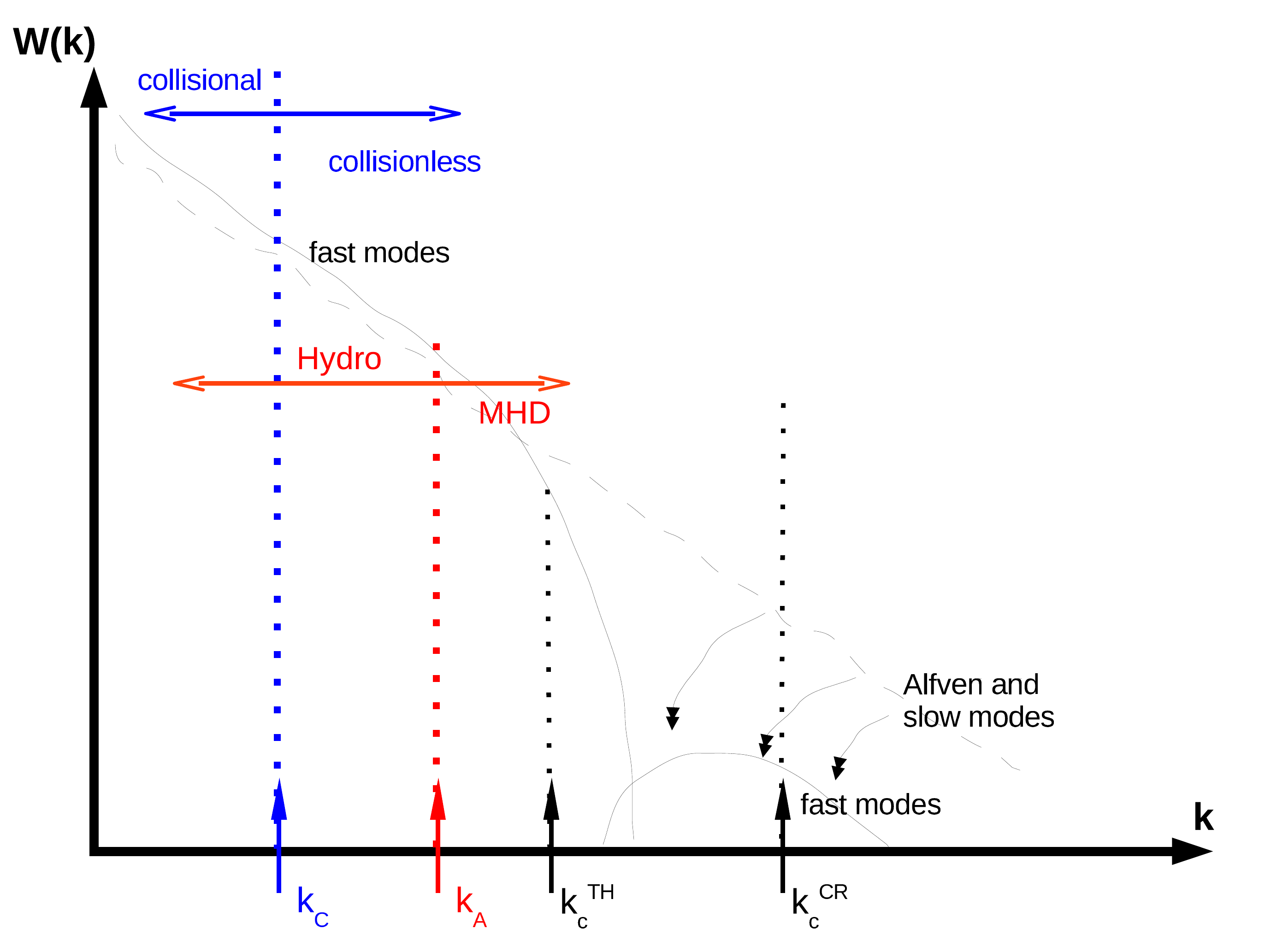}
\includegraphics[width=0.417\textwidth]{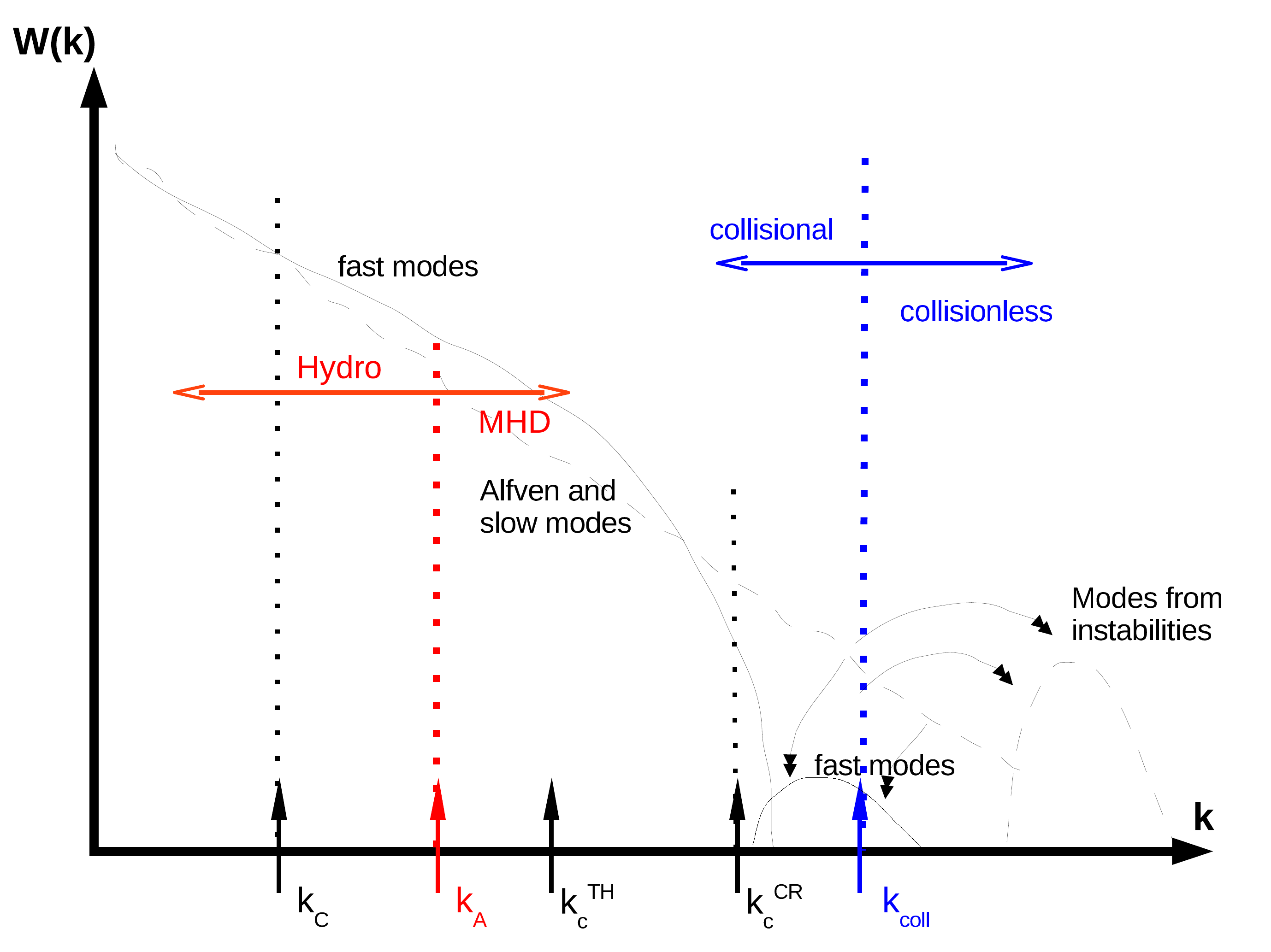}
\caption[]{A cartoon giving the evolution with scale of turbulence 
(solid lines $=$ fast modes, dashed lines $=$ Alfv\'en and slow modes) 
assuming collisionless interaction with thermal IGM (e.g. 
Brunetti \& Lazarian 2007) ({\bf Left Panel}) and collisional 
interaction mediated by plasma instabilities (this paper) 
({\bf Right Panel}).
In the Figure we also mark the generation (thin arrows)
of fast modes from the cascading 
of Alfv\'en and slow modes (both panels) and that of solenoidal modes 
from plasma instabilities (right panel).
The position of the relevant scales, Coulomb scale ($k_C$), 
MHD scale ($l_A^{-1}\sim k_A$), thermal and relativistic cut--off 
scales ($k_c^{th}$ and $k_c^{CR}$, respectively) is reported, 
together with that of the effective collisionless scale $k_{coll}$ 
(in the right panel; $k_{coll}=k_{C}$ in the left panel). 
The exponential cut--off in the spectrum of fast modes marks the effect 
of collisionless damping with thermal IGM (left panel) and with 
relativistic particles (right panel).} 
\label{fig:substructures}
\end{figure*}

\noindent
The typical spatial scale of the collisionless (TTD) thermal
damping is fairly large,
$1/k_c^{th} \sim 0.1-1 \times {{L_o(kpc)}\over{300}}
( {{V_L^2/c_s^2}\over{0.3}} )^{-2}$ kpc (Brunetti \& Lazarian 2007).
Consequently it is reasonable to assume that plasma instabilities driven
by compressible turbulence (even considering a modest level of
turbulence) may maintain the IGM at least {\it weakly--collisionless},
with $k_c^{CR} > k_{coll} >> k_c^{th}$,
allowing for an increasing fraction of turbulence to become available
for reacceleration of relativistic particles with respect to the
case of {\it collisionless} IGM.
In this regime the damping of
turbulence is collisional (as in Sect.~4) at
scales larger than $\sim 1/k_{coll}$, 
whereas the cascade of fast modes
is suddenly interrupted by collisionless TTD with thermal electrons
as soon as $1/k \leq 1/k_{coll}$.
In this regime the momentum--diffusion coefficient 
of relativistic particles subject to TTD resonance can be estimated
as (from Eq.(47) in Brunetti \& Lazarian 2007)\footnote{Here we
are assuming that the dominant damping of fast modes is still
provided by
the resonance with thermal particles at scales $\leq 1/k_{coll}$} :

\begin{equation}
D_{pp} \approx 
{{\pi}\over{8}} \, {{p^2}\over{c}}
\langle {{ \beta_{pl} |B_k|^2 }\over{16 \pi {{\cal W}}^f}} \rangle
\left( {{ 2 I_o^f c_s}\over{7 \rho}} \right)^{ {1\over2} }
{ k_{coll} }^{ {1\over2} } 
{\cal I}
\label{dpp_ICM}
\end{equation}

\noindent
where both ${\cal I}$ and $\langle ... \rangle$ $\sim$ a few.

\noindent
As in Sect.~2 we consider the case of firehose instability
where the collisionless scale is (from
Eqs.\ref{firehose} and \ref{kd}):

\begin{equation}
k_{coll} \sim 8 \, k_o f_{slow}^{3/2} \left( {{V_L}\over{c_s}} \right)^3
\left( {{c_s}\over{v_A}} \right)^9
\label{kcoll2}
\end{equation}

\noindent
where $f_{slow}$ is the ratio of the energy of slow and 
fast modes. 
Consequently the momentum--diffusion coefficient in the
{\it weakly--collisionless} regime can be estimated :

\begin{equation}
D_{pp} \sim {{\cal C}}_D \, p^2 c_s^2 k_o
\left( {{V_L}\over{c_s}} \right)^{7 \over 2}
\left( {{c_s}\over{v_A}} \right)^{9 \over 2}
f_{slow}^{3/4} \,\,\, ,
\label{DppWC}
\end{equation}

\noindent 
(${{\cal C}}_D$ a constant) 
i.e. $D_{pp} \propto T \beta_{pl}^{9/4}$ (for a fixed $V_L / c_s$), that 
has the same scaling with temperature derived in the collisionless 
regime.

As explained in the previous Section, an increasing fraction of 
turbulent energy available for the reacceleration of cosmic rays 
does not imply that the acceleration efficiencies are
much larger than in the {\it collisionless} 
case, since if we assume that 
most of the turbulent energy is channelled into
cosmic rays the back reaction of these particles 
self-regulates the acceleration process
approaching a universal regime.

Non--thermal radiations from galaxy clusters 
are probes of in situ particle acceleration processes in the IGM
that are activated in massive (hot) galaxy clusters
during cluster--cluster mergers (e.g. Cassano et al 2010).
Previous turbulent reacceleration models for the origin
of giant radio halos were based on the assumption of 
a {\it collisionless} IGM.
In the context of these models the fact that nowadays
giant radio halos are found only in massive (and hot) clusters is 
interpreted via simple energetics arguments that stem in the 
expectation that the turbulent injection rate increases with the mass 
of the merging clusters (e.g. Cassano \& Brunetti 2005).

\noindent
Additional inputs may come from our explorative study.
We believe that the picture may be more complex than previously 
thought. 
Assuming that the compressible turbulence is generated at large
(injection) scales and then cascades to smaller scales,
we suggest that the effects of reduced mean free path in
a turbulent IGM may allow the {\it fraction} 
of the energy of turbulence that is available to the reacceleration of
relativistic particles to be larger than that derived for
a {\it collisionless} IGM. This readily implies  
that the damping from relativistic particles plays a role
in regulating the acceleration efficiency and introduces a
new physical threshold in the mechanism responsible for the reacceleration 
of the relativistic particles.
The threshold is connected with the energy contributed 
by cosmic rays in the IGM: 
the level of compressible turbulence in the IGM of hot, merging,
clusters may become comparable to (or larger than) that 
contributed by cosmic rays, while the damping due to
these particles may suppress particle reacceleration in less
turbulent, relaxed clusters.

\noindent
On the other hand this complex picture does not affect the basic
expectations from previous studies.
Indeed we show that the scaling of the acceleration
efficiency with IGM temperature derived assuming a
{\it collisionless} IGM 
may also extend to the case of a {\it weakly--collisionless} IGM
implying that the conclusion that
stochastic acceleration is stronger in the hottest clusters
holds for a wide range of (micro--)physical conditions.
By considering the example where the collisionless
scale is regulated by the effects of firehose instability,
both the acceleration efficiency and the collisional
or collisionless nature of the interaction between turbulence and
the IGM depend on the beta of the plasma plays, 
with the IGM being more {\it collisional}
for larger values of $\beta_{pl}$ (Eq.\ref{kcoll2}).

In this paper we have considered 
only the case of stochastic reacceleration
due to compressible turbulence (essentially fast modes).
However we do not think that the overall picture of turbulent
reacceleration described in the present paper is complete.
Indeed the processes of {\it in situ}
reacceleration of particles in the clusters of galaxies
is not limited to the acceleration by fast modes. For instance, processes
of magnetic reconnection in turbulent astrophysical plasmas
can induce additional acceleration (see Lazarian \&
Opher 2009, Drake et al. 2010, Lazarian \& Desiati 2010, Brunetti \& Lazarian
2010 and references therein).
Also Alfv\'en modes were proposed for the reacceleration of
relativistic particles in the IGM (e.g. Fujita et al 2003,
Brunetti et al 2004). For instance instabilities in cosmic ray
collisionless fluid can induce small-scales
Alfven waves with ${\bf k}$ parallel to
magnetic fields and those waves can efficiently induce cosmic ray
reacceleration (Lazarian \& Beresnyak 2006, Brunetti \& Lazarian 2010, 
Yan \& Lazarian in prep.), a process that requires further investigation.
We want to stress that overall,
the existence of all these processes in a turbulent IGM
is suggestive that even larger share of
energy (compared to our present estimates) of turbulence energy 
can be transferred to relativistic particles in the galaxy clusters.

\section{Conclusions}
\label{sec:conclu}

It has been proposed that the observed giant radio halos may be due to
turbulent reacceleration of relativistic particles in merging
clusters.
Calculations of particle acceleration by MHD turbulence must 
account self-consistently for the non--linear interaction 
between turbulent waves and particles.
Previous theoretical works in this context 
focus on the interaction between compressible
turbulence (fast modes) and both the 
thermal IGM and the relativistic particles by assuming 
a {\it collisionless} IGM. In this case about 90 percent of the
energy of fast modes goes into heating of the thermal plasma, while only 
10 percent is available for the reacceleration of relativistic 
particles (eg. Brunetti \& Lazarian 2007). 

In this paper we explore a new possibility for the 
particle reacceleration by compressible
turbulence where the interaction between
turbulent modes and the thermal IGM is {\it collisional} 
at scales much smaller than the Coulomb mean free path.
We have motivated this guess by observing that several plasma
instabilities can be generated by turbulent motions in the
IGM driving perturbations in the magnetic field that induce scattering
of charged particles on scales much smaller than the Coulomb
mean free path.
This process results in a {\it collective} interaction of individual 
ions with the rest of the plasma and potentially
constrains the effective collisionless scale 
to (about) the scales where instabilities develop.

\noindent
Under these conditions we find that an increasing fraction of the
energy of fast modes is available to the reacceleration of
relativistic particles and that the collisionless
damping of turbulent motions induced by these particles contribute to
self-regulate the acceleration process.
Assuming a {\it collisional} IGM, all the energy of fast modes
is channeled into the reacceleration of relativistic particles.
Interestingly in this case the upper limits to the energy densities
of cosmic rays in galaxy clusters derived by recent gamma and radio
observations provide also constraints to the
fraction of the thermal energy in the IGM available for fast modes.
In a {\it collisional} IGM the 
acceleration efficiency results from the balance between
the injection rate of compressible turbulent motions and the rate of
turbulent damping due to the reaccelerated particles.
In this case we find that a universal acceleration regime is established
as soon as turbulent energy approaches equipartition with the energy
density of relativistic particles, and that in this case typical
(systematic) 
reacceleration times $\sim 10^8$ yrs are provided by TTD with fast
modes.

\noindent
Under the assumption of a {\it collisional} IGM we calculate the
reacceleration of relativistic protons and of secondary 
relativistic electrons.
Based on present constraints from FERMI observations of nearby
clusters, we consider a situation where cosmic rays contribute to a few
percent of the thermal energy of the IGM and show that relativistic
electrons can be reaccelerated at energies of several GeV, provided that
compressible turbulence, generated at large scales, contributes to several 
percent of the cluster thermal energy.
The reaccelerated electrons emit synchrotron radiation at GHz frequencies
in typical (several) $\mu$G magnetic fields providing an explanation for 
the origin of the observed giant radio halos.
Remarkably, in the case of {\it collisional} IGM the amount of cluster 
turbulence necessary to reaccelerate
GeV electrons is significantly smaller than that 
in models that assume a {\it collisionless} IGM.

\noindent
Whether the interaction between turbulence and IGM behaves 
{\it collisionless} or {\it collisional} depends on the effect 
of the reduced mean free path.
It is reasonable to assume that the IGM becomes 
(at least) {\it weakly--collisionless} as soon as
a sufficient (even modest) level of compressible turbulence is generated.
This has important consequences on the process of particle acceleration
by compressible turbulence in the IGM.
All the calculations of turbulent reacceleration developed
under the assumptions of a {\it collisionless} IGM
should be retained as conservative approaches.
Consequently our explorative study provides further
theoretical support to the idea that turbulence may play an important 
role in the the origin of non-thermal components (and emission) 
on cluster scales.

Finally we want to remark that although we believe that fast modes may
play an important role in the process of {\it in situ} reacceleration, 
additional processes in a turbulent IGM (e.g. reconnection, 
resonance with small scale Alfv\'en waves, etc) may allow a larger fraction of 
the energy of cluster--turbulence to become available for 
the reacceleration of relativistic particles.
Further research should quantify these processes.

\section{Acknowledgements}
We thank G.Setti for useful comments.
We acknowledge grants from INAF (PRIN-INAF2007 and 2008),
ASI-INAF (I/088/06/0), NSF (AST 0808118) and NASA (NNX09AH78G)
and support by the NSF Center for Magnetic Self-Organization.
GB thanks the Dep. of Astronomy of Wisconsin University at Madison
and the Harvard-Smithsonian Center for Astrophysics for hospitality.

\end{document}